\documentclass[preprint,superscriptaddress,aps,prl,superscriptaddress,showpacs,preprintnumbers,amsmath,amssymb]{revtex4}
\usepackage{graphicx}
\usepackage{xspace}
\usepackage{epsfig}
\usepackage{epstopdf}
\usepackage{multirow}
\usepackage{dcolumn}  
\usepackage{wasysym}
\usepackage[section]{placeins}
\usepackage{lineno}
\usepackage{subfigure}
\usepackage{color}

\graphicspath{{ps}}

\begin{document}

\preprint{\vbox{ \hbox{   }
							   \hbox{Belle Preprint 2016-13}
							   \hbox{KEK Preprint 2016-51}
}}

\title{ \quad\\[1.0cm]\Large \bf \boldmath Search for $D^{0}$ decays to invisible final states at Belle}


\begin{abstract}
We report the result from the first search for $D^0$ decays to invisible final states. The analysis is performed on a data sample of 924 $\rm{fb}^{-1}$ collected at and near the $\Upsilon(4S)$ and $\Upsilon(5S)$ resonances with the Belle detector at the KEKB asymmetric-energy $e^{+}e^{-}$ collider. The absolute branching fraction is determined using an inclusive $D^0$ sample, obtained by fully reconstructing the rest of the particle system including the other charmed particle. No significant signal yield is observed and an upper limit of $9.4\times 10^{-5}$ is set on the branching fraction of $D^0$ to invisible final states at 90\% confidence level.

\end{abstract}

\affiliation{University of the Basque Country UPV/EHU, 48080 Bilbao}
\affiliation{Beihang University, Beijing 100191}
\affiliation{Budker Institute of Nuclear Physics SB RAS, Novosibirsk 630090}
\affiliation{Faculty of Mathematics and Physics, Charles University, 121 16 Prague}
\affiliation{Chonnam National University, Kwangju 660-701}
\affiliation{University of Cincinnati, Cincinnati, Ohio 45221}
\affiliation{Deutsches Elektronen--Synchrotron, 22607 Hamburg}
\affiliation{University of Florida, Gainesville, Florida 32611}
\affiliation{Justus-Liebig-Universit\"at Gie\ss{}en, 35392 Gie\ss{}en}
\affiliation{SOKENDAI (The Graduate University for Advanced Studies), Hayama 240-0193}
\affiliation{Hanyang University, Seoul 133-791}
\affiliation{University of Hawaii, Honolulu, Hawaii 96822}
\affiliation{High Energy Accelerator Research Organization (KEK), Tsukuba 305-0801}
\affiliation{J-PARC Branch, KEK Theory Center, High Energy Accelerator Research Organization (KEK), Tsukuba 305-0801}
\affiliation{IKERBASQUE, Basque Foundation for Science, 48013 Bilbao}
\affiliation{Indian Institute of Science Education and Research Mohali, SAS Nagar, 140306}
\affiliation{Indian Institute of Technology Bhubaneswar, Satya Nagar 751007}
\affiliation{Indian Institute of Technology Guwahati, Assam 781039}
\affiliation{Indian Institute of Technology Madras, Chennai 600036}
\affiliation{Indiana University, Bloomington, Indiana 47408}
\affiliation{Institute of High Energy Physics, Chinese Academy of Sciences, Beijing 100049}
\affiliation{Institute of High Energy Physics, Vienna 1050}
\affiliation{INFN - Sezione di Torino, 10125 Torino}
\affiliation{J. Stefan Institute, 1000 Ljubljana}
\affiliation{Kanagawa University, Yokohama 221-8686}
\affiliation{Institut f\"ur Experimentelle Kernphysik, Karlsruher Institut f\"ur Technologie, 76131 Karlsruhe}
\affiliation{Kennesaw State University, Kennesaw, Georgia 30144}
\affiliation{King Abdulaziz City for Science and Technology, Riyadh 11442}
\affiliation{Department of Physics, Faculty of Science, King Abdulaziz University, Jeddah 21589}
\affiliation{Korea Institute of Science and Technology Information, Daejeon 305-806}
\affiliation{Korea University, Seoul 136-713}
\affiliation{Kyungpook National University, Daegu 702-701}
\affiliation{\'Ecole Polytechnique F\'ed\'erale de Lausanne (EPFL), Lausanne 1015}
\affiliation{P.N. Lebedev Physical Institute of the Russian Academy of Sciences, Moscow 119991}
\affiliation{Faculty of Mathematics and Physics, University of Ljubljana, 1000 Ljubljana}
\affiliation{Ludwig Maximilians University, 80539 Munich}
\affiliation{Luther College, Decorah, Iowa 52101}
\affiliation{University of Maribor, 2000 Maribor}
\affiliation{Max-Planck-Institut f\"ur Physik, 80805 M\"unchen}
\affiliation{School of Physics, University of Melbourne, Victoria 3010}
\affiliation{University of Miyazaki, Miyazaki 889-2192}
\affiliation{Moscow Physical Engineering Institute, Moscow 115409}
\affiliation{Moscow Institute of Physics and Technology, Moscow Region 141700}
\affiliation{Graduate School of Science, Nagoya University, Nagoya 464-8602}
\affiliation{Kobayashi-Maskawa Institute, Nagoya University, Nagoya 464-8602}
\affiliation{Nara Women's University, Nara 630-8506}
\affiliation{National Central University, Chung-li 32054}
\affiliation{National United University, Miao Li 36003}
\affiliation{Department of Physics, National Taiwan University, Taipei 10617}
\affiliation{H. Niewodniczanski Institute of Nuclear Physics, Krakow 31-342}
\affiliation{Nippon Dental University, Niigata 951-8580}
\affiliation{Niigata University, Niigata 950-2181}
\affiliation{Novosibirsk State University, Novosibirsk 630090}
\affiliation{Osaka City University, Osaka 558-8585}
\affiliation{Pacific Northwest National Laboratory, Richland, Washington 99352}
\affiliation{University of Pittsburgh, Pittsburgh, Pennsylvania 15260}
\affiliation{Theoretical Research Division, Nishina Center, RIKEN, Saitama 351-0198}
\affiliation{University of Science and Technology of China, Hefei 230026}
\affiliation{Showa Pharmaceutical University, Tokyo 194-8543}
\affiliation{Soongsil University, Seoul 156-743}
\affiliation{Stefan Meyer Institute for Subatomic Physics, Vienna 1090}
\affiliation{Sungkyunkwan University, Suwon 440-746}
\affiliation{School of Physics, University of Sydney, New South Wales 2006}
\affiliation{Department of Physics, Faculty of Science, University of Tabuk, Tabuk 71451}
\affiliation{Tata Institute of Fundamental Research, Mumbai 400005}
\affiliation{Excellence Cluster Universe, Technische Universit\"at M\"unchen, 85748 Garching}
\affiliation{Department of Physics, Technische Universit\"at M\"unchen, 85748 Garching}
\affiliation{Toho University, Funabashi 274-8510}
\affiliation{Department of Physics, Tohoku University, Sendai 980-8578}
\affiliation{Earthquake Research Institute, University of Tokyo, Tokyo 113-0032}
\affiliation{Department of Physics, University of Tokyo, Tokyo 113-0033}
\affiliation{Tokyo Institute of Technology, Tokyo 152-8550}
\affiliation{Tokyo Metropolitan University, Tokyo 192-0397}
\affiliation{University of Torino, 10124 Torino}
\affiliation{Utkal University, Bhubaneswar 751004}
\affiliation{Virginia Polytechnic Institute and State University, Blacksburg, Virginia 24061}
\affiliation{Wayne State University, Detroit, Michigan 48202}
\affiliation{Yamagata University, Yamagata 990-8560}
\affiliation{Yonsei University, Seoul 120-749}
  \author{Y.-T.~Lai}\affiliation{Department of Physics, National Taiwan University, Taipei 10617}
  \author{M.-Z.~Wang}\affiliation{Department of Physics, National Taiwan University, Taipei 10617}

  \author{I.~Adachi}\affiliation{High Energy Accelerator Research Organization (KEK), Tsukuba 305-0801}\affiliation{SOKENDAI (The Graduate University for Advanced Studies), Hayama 240-0193} 
  \author{H.~Aihara}\affiliation{Department of Physics, University of Tokyo, Tokyo 113-0033} 
  \author{S.~Al~Said}\affiliation{Department of Physics, Faculty of Science, University of Tabuk, Tabuk 71451}\affiliation{Department of Physics, Faculty of Science, King Abdulaziz University, Jeddah 21589} 
  \author{D.~M.~Asner}\affiliation{Pacific Northwest National Laboratory, Richland, Washington 99352} 
  \author{T.~Aushev}\affiliation{Moscow Institute of Physics and Technology, Moscow Region 141700} 
  \author{R.~Ayad}\affiliation{Department of Physics, Faculty of Science, University of Tabuk, Tabuk 71451} 
  \author{I.~Badhrees}\affiliation{Department of Physics, Faculty of Science, University of Tabuk, Tabuk 71451}\affiliation{King Abdulaziz City for Science and Technology, Riyadh 11442} 
  \author{A.~M.~Bakich}\affiliation{School of Physics, University of Sydney, New South Wales 2006} 
% ...
\collaboration{The Belle Collaboration}
\noaffiliation

\maketitle
In the Standard Model (SM), heavy ($D$ or $B$) meson decay to $\nu\overline{\nu}$ is helicity suppressed~\cite{dm} with an expected branching fraction of $\mathcal{B}(D^{0}\to\nu\overline{\nu})=1.1\times10^{-30}$~\cite{charge_conjugate}, which is beyond the reach of current collider experiments. 
The branching fraction may be enhanced by non-SM mechanisms such as the decay of $D$ and $B$ mesons to dark matter (DM) final states with and without an additional light meson in the final states, as estimated in Ref.~\cite{dm}.
With several DM candidates~\cite{vquark,right_nu}, the branching fraction of $D^{0}$ to invisible final states could be enhanced to $\mathcal{O}(10^{-15})$. 

Recent DM searches are mainly based on the direct detection of the nuclear recoil signal due to DM interaction~\cite{LUX,CDMS}, or $\gamma$-ray, $e^{+}e^{-}$ and $p\overline{p}$ production due to DM annihilation~\cite{FERMI,PAMELA}. At an $e^{+}e^{-}$ ``flavor factory,'' in which two heavy-flavor particles are produced in flavor-conjugate states, the indirect detection of DM candidates is performed as follows. One of the $D$ or $B$ mesons is fully reconstructed, and then energy-momentum conservation is used to search for the decay of the other $D$ or $B$ meson into an invisible final state.

\begin{figure}[htb]
\centering
\includegraphics[height=5cm]{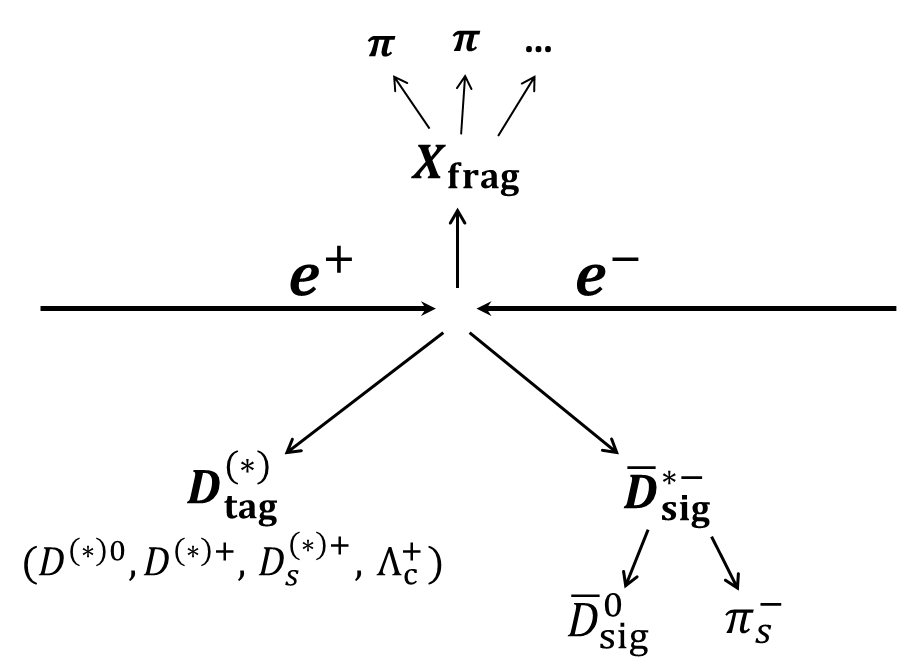}
\caption{An illustration of the charm tagger method.}
\label{fg:charm_tagger} 
\end{figure}

In Belle, a few hundred million $D$ mesons are produced in $e^{+}e^{-}\to c\overline{c}$ continuum events. We use the charm tagger method to select an inclusive $D^{0}$ sample, which permits the identification of $D^{0}$ decays involving invisible particles~\cite{widhalm0, widhalm1, anze, babar_tagger}: the process $e^{+}e^{-}\to c\overline{c} \to D^{(*)}_{\textrm{tag}}X_{\textrm{frag}}\overline{D}^{*-}_{\textrm{sig}}$ with $\overline{D}^{*-}_{\textrm{sig}}\to \overline{D}^{0}_{\textrm{sig}}\pi^{-}_{s}$ is reconstructed except for $\overline{D}^{0}_{\textrm{sig}}$, as illustrated in Fig.~\ref{fg:charm_tagger}. Here, $D^{(*)}_{\textrm{tag}}$ represents a charmed particle used as a tag: $D^{(*)0}$, $D^{(*)+}$, $D^{(*)+}_{s}$, or $\Lambda^{+}_{c}$. Since the center-of-mass (c.m.) energy of KEKB is above the open charm threshold, a fragmentation system ($X_{\textrm{frag}}$) with a few light unflavored mesons may also be produced. The $\pi^{-}_{s}$ denotes a charged pion from $\overline{D}^{*-}_{\textrm{sig}}$ decay.

This search for $D^{0}\to$ invisible decay with the charm tagger method at $B$ factories provides a powerful way to search for DM: any clear signal would be an indication for new physics. Measurements of $B^{0}\to$ invisible with both hadronic and semileptonic $B$ tagging methods are already reported by both Belle and {\it BABAR}~\cite{btonunu, btonunu_1}.

We use the data sample of 924 $\rm{fb^{-1}}$ collected at or near the $\Upsilon(4S)$ and $\Upsilon(5S)$ resonances with the Belle detector~\cite{Belle} at the KEKB asymmetric-energy $e^+ e^-$ collider~\cite{KEKB}. 
The Belle detector is a large-solid-angle magnetic spectrometer that consists of a silicon vertex detector (SVD), a 50-layer central drift chamber (CDC), an array of aerogel threshold Cherenkov counters (ACC), a barrel-like arrangement of time-of-flight scintillation counters (TOF) and an electromagnetic calorimeter (ECL) composed of CsI(Tl) crystals located inside a superconducting solenoid that provides a 1.5~T magnetic field. An iron flux-return yoke located outside the solenoid is instrumented to detect $K_L^0$ mesons and to identify muons. 

This analysis uses the data sets with two different inner-detector configurations.
About 156 $\rm{fb}^{-1}$ were collected with a beam pipe of radius 2 cm and with three layers of SVD, while the rest of the data set was collected with a beam pipe of radius 1.5~cm and four layers of SVD \cite{svd2}. 
Large Monte Carlo (MC) samples for signal and several backgrounds are generated with EvtGen~\cite{ref:EvtGen} and simulated with GEANT3~\cite{geant} with the configurations of the Belle detector. These samples are used to obtain expected distributions of various physical quantities for signal and background, to optimize the selection criteria, and to determine the signal selection efficiency. 

We use the knowledge of the $e^{+}e^{-}$ four-momentum to identify a $D^{0}$ that escaped detection by fully reconstructing the remainder of the event (whether this $D^{0}$ decays visibly or not).
The four types of $D_{\textrm{tag}}$ are reconstructed using 23 decay modes. ($D^{*}_{\textrm{tag}}$ candidates are described later.) The decay modes and the corresponding requirements on the $D_{\textrm{tag}}$ momentum in the c.m.\ frame ($p^{*}$) are listed in Table~\ref{tb:d_tag}; these requirements were optimized in Ref.~\cite{anze}.

\begin{table}[htbp]
\begin{center}
\caption{$D_{\textrm{tag}}$ decay modes and corresponding requirements on the $D_{\textrm{tag}}$ momentum in the c.m. frame ($p^{*}$).}
\begin{tabular}{l|c||l|c}
\hline\hline
$D^{0}$ decay & $p^{*}$ (GeV/$c$) & $D^{+}$ decay & $p^{*}$ (GeV/$c$)  \\
\hline
$K^{-}\pi^{+}$ & $> 2.3$ & $K^{-}\pi^{+}\pi^{+}$ & $> 2.3$\\
$K^{-}\pi^{+}\pi^{0}$ & $> 2.5$ & $K^{-}\pi^{+}\pi^{+}\pi^{0}$ & $> 2.5$\\
$K^{-}\pi^{-}\pi^{+}\pi^{+}$ & $> 2.3$ & $K^{0}_{S}\pi^{+}$ & $> 2.3$\\
$K^{-}\pi^{-}\pi^{+}\pi^{+}\pi^{0}$ & $> 2.5$ & $K^{0}_{S}\pi^{+}\pi^{0}$ & $> 2.4$\\
$K^{0}_{S}\pi^{+}\pi^{-}$ & $> 2.3$ & $K^{0}_{S}\pi^{+}\pi^{+}\pi^{-}$ & $> 2.4$\\
$K^{0}_{S}\pi^{+}\pi^{-}\pi^{0}$ & $> 2.5$ & $K^{+}K^{-}\pi^{+}$ & $> 2.3$\\
\hline\hline
\multicolumn{4}{c}{} \\
\hline\hline
$\Lambda^{+}_{c}$ decay & $p^{*}$ (GeV/$c$) & $D^{+}_{s}$ decay & $p^{*}$ (GeV/$c$) \\
\hline
$pK^{-}\pi^{+}$ & $> 2.3$  & $K^{+}K^{-}\pi^{+}$ & $> 2.3$ \\
$pK^{-}\pi^{+}\pi^{0}$ & $> 2.5$ & $K^{0}_{S}K^{+}$ & $> 2.3$ \\
$pK^{0}_{S}$ & $> 2.3$ & $K^{0}_{S}K^{0}_{S}\pi^{+}$ & $> 2.3$ \\
$\Lambda\pi^{+}$ & $> 2.3$ & $K^{+}K^{-}\pi^{+}\pi^{0}$ & $> 2.5$\\
$\Lambda\pi^{+}\pi^{0}$ & $> 2.5$ & $K^{0}_{S}K^{-}\pi^{+}\pi^{+}$ & $> 2.4$\\
$\Lambda\pi^{+}\pi^{+}\pi^{-}$ & $> 2.3$ & &\\
\hline\hline
\end{tabular}
\label{tb:d_tag}
\end{center}
\end{table}

The selection criteria for the final-state charged particles in $D_{\textrm{tag}}$ are based on information obtained from the tracking systems (SVD and CDC) and the hadron identification systems (CDC, ACC, and TOF). These particles are required to have an impact parameter within $\pm 0.5$ cm of the interaction point (IP) in the transverse plane, and within $\pm 1.5$ cm along the positron beam direction.
The likelihood values of each track for different particle types, $L_{p}$, $L_{K}$, and $L_{\pi}$, are determined from the information provided by the hadron-identification system. The track is identified as a proton if $L_K/(L_K+L_p) < 0.9$ and $L_{\pi}/(L_{\pi}+L_p) < 0.9$, as a pion if $L_K/(L_K+L_{\pi}) < 0.9$, and as a kaon if $L_K/(L_K+L_{\pi}) > 0.1$. The efficiencies are about 99\% for identifying each type of charged hadron.

Photons are reconstructed from the energy clusters in the ECL that are not associated with charged tracks. A $\pi^{0}$ is reconstructed from two photon candidates by requiring the di-photon invariant mass ($M_{\gamma\gamma}$) to be between 0.115 and 0.150 GeV/$c^{2}$ (with an efficiency of 89\%). The energy of each photon candidate is required to be greater than 50 MeV and a mass-constrained fit is performed on the reconstructed $\pi^{0}$ candidate.
For the $D_{\textrm{tag}}$ channels with more than two tracks, a $K^{0}_{S}$ and two tracks, or a $\Lambda$ in the final states, the photons are required to have an energy greater than 100 MeV in the ECL endcaps. 

The $K^{0}_{S}$ ($\Lambda$) candidates are reconstructed in the $\pi^{+}\pi^{-}$ ($p\pi^{-}$) mode and are required to have invariant $M_{\pi^{+}\pi^{-}}$ ($M_{p\pi^{-}}$) between 0.468 and 0.508 GeV/$c^{2}$ (1.111 and 1.121 GeV/$c^{2}$), leading to an efficiency of about 64\% (47\%). A successful vertex fit is also required ($\chi^{2}<100$ for $\Lambda$). The $K^{0}_{L}$ candidates are reconstructed from the clusters in KLM that are not associated with charged tracks.

The $D_{\textrm{tag}}$ candidates are required to have an invariant mass within $\pm3 \sigma$ of the nominal mass~\cite{PDG}~(where $\sigma$ is the resolution of measurement) and be successfully fit to a common vertex with a mass constraint.

The $D^{*}_{\textrm{tag}}$ candidates are reconstructed via five decay modes: $D^{*+}\to D^{0}\pi^{+}$, $D^{*+}\to D^{+}\pi^{0}$, $D^{*0}\to D^{0}\pi^{0}$, $D^{*0}\to D^{0}\gamma$, and $D^{*+}_{s}\to D^{+}_{s}\gamma$. The $\gamma$ candidate used in $D^{*0}$ or $D^{*+}_{s}$ reconstructions is required to have an energy greater than 0.12 GeV and is paired with all other photons in the event to ensure that it is not from a $\pi^{0}$ decay: if $M_{\gamma\gamma}$ is within $\pm$10 MeV/$c^{2}$ of the nominal $\pi^{0}$ mass and the energy asymmetry ($|(E_{\gamma1}-E_{\gamma2})/(E_{\gamma1}+E_{\gamma2})|$) is less than 0.5, the $D^{*0}$ or $D^{*+}_{s}$ candidate is rejected. The mass difference between the $D^{*}_{\textrm{tag}}$ and $D_{\textrm{tag}}$ is required to be within $\pm3 \sigma$ of the nominal $D^{*}_{(s)}-D_{(s)}$ mass difference~\cite{PDG}. The $\pi^{+}$ from the $D^{*}_{\textrm{tag}}$ decay is refitted to the $D_{\textrm{tag}}$ vertex.

The $X_{\textrm{frag}}$ system is reconstructed from the remaining particles as listed in Table~\ref{tb:x_frag}. The charge of $D^{(*)}_{\textrm{tag}}X_{\textrm{frag}}$ is required to be $+1$~\cite{charge_conjugate}. For each combination of $D^{(*)}_{\textrm{tag}}$ $X_{\textrm{frag}}$, the missing mass recoiling against $D^{(*)}_{\textrm{tag}}$ $X_{\textrm{frag}}$, $M_{\textrm{miss}}(D^{(*)}_{\textrm{tag}}X_{\textrm{frag}})$, is required to be between 1.86 and 2.16 GeV/$c^{2}$ to select a $\overline{D}^{*-}_{\textrm{sig}}$ candidate. At this stage, all candidates satisfying the selection criteria are retained.

For each $D^{(*)}_{\textrm{tag}}$$X_{\textrm{frag}}$ candidate satisfying the above $M_{\textrm{miss}}(D^{(*)}_{\textrm{tag}}X_{\textrm{frag}})$ requirement, the remaining tracks not associated with $D^{(*)}_{\textrm{tag}}$$X_{\textrm{frag}}$ are examined for a $\pi^{-}_{s}$ candidate. For each such candidate, the missing momentum recoiling against the $D^{(*)}_{\textrm{tag}}$ $X_{\textrm{frag}}$ $\pi^{-}_{s}$ system in the c.m.\ frame is calculated and required to be greater than 2.0 GeV/$c$. The missing mass for the $D^{(*)}_{\textrm{tag}}$ $X_{\textrm{frag}}$ $\pi^{-}_{s}$ system ($M_{D^{0}}$) is subsequently calculated from a fit in which $M_{\textrm{miss}}(D^{(*)}_{\textrm{tag}}X_{\textrm{frag}})$ is constrained to the nominal $D^{*+}$ mass ($m_{D^{*+}}$)~\cite{PDG}~(to improve the resolution). 
If more than one $\overline{D}^{0}_{\textrm{sig}}$ candidate is found in an event, we first choose the one with the smallest $\chi^{2}$, which is obtained from the fit with $M_{\textrm{miss}}(D^{(*)}_{\textrm{tag}}X_{\textrm{frag}})$ constrained to $m_{D^{*+}}$. If still more than one candidate is found (with multiple $\pi_{s}$'s), we choose the one with the largest opening angle between $\overline{D}^{0}_{\textrm{sig}}$ and $D^{(*)}_{\textrm{tag}}$ in the c.m.\ frame. 
Multiple candidates are found in 56.6\% of the data with an average multiplicity of inclusive $D^{0}$ candidates of 2.7, which is consistent with MC simulation.

\begin{table}[htbp]
\begin{center}
\caption{$X_{\textrm{frag}}$ system for $D^{(*)}_{\textrm{tag}}$.}
\begin{tabular}{c||cc} 
\hline\hline
$D^{(*)+}$ & \multicolumn{2}{c}{$D^{(*)0}$} \\
\hline
nothing($K^{+}K^{-}$) & \multicolumn{2}{c}{$\pi^{+}(K^{+}K^{-})$} \\
$\pi^{0}(K^{+}K^{-})$ & \multicolumn{2}{c}{$\pi^{+}\pi^{0}(K^{+}K^{-})$} \\
$\pi^{+}\pi^{-}(K^{+}K^{-})$ & \multicolumn{2}{c}{$\pi^{+}\pi^{-}\pi^{+}(K^{+}K^{-})$} \\
$\pi^{+}\pi^{-}\pi^{0}(K^{+}K^{-})$ & &  \\
\hline\hline
\multicolumn{2}{c}{} \\
\hline\hline
$\Lambda^{+}_{c}$ & \multicolumn{2}{c}{$D^{(*)+}_{s}$} \\
\hline
$\pi^{+}\overline{p}$ & $K^{0}_{S}$, & $\pi^{0}K^{0}_{S}$ \\
$\pi^{+}\pi^{0}\overline{p}$ & $\pi^{+}K^{-}$, & $\pi^{+}\pi^{0}K^{-}$ \\
$\pi^{+}\pi^{-}\pi^{+}\overline{p}$ & $\pi^{+}\pi^{-}K^{0}_{S}$, & $\pi^{+}\pi^{-}\pi^{0}K^{0}_{S}$ \\
& $\pi^{+}\pi^{-}\pi^{+}K^{-}$ &  \\
\hline\hline
\end{tabular}
\label{tb:x_frag}
\end{center}
\end{table}

\begin{figure}[htb]
\centering
\includegraphics[height=0.4\textwidth]{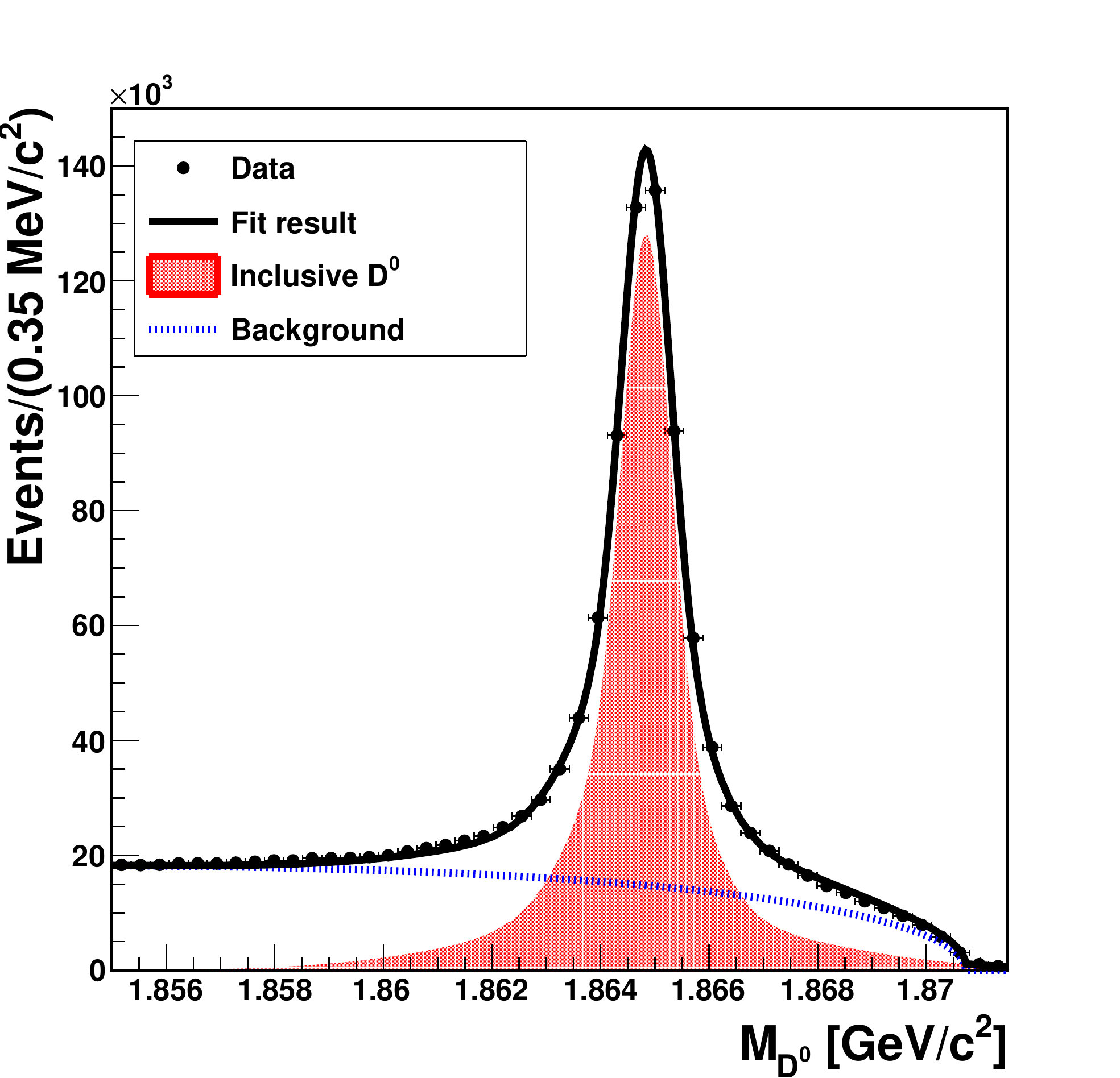}
\caption{The $M_{D^{0}}$ distribution of the inclusive $D^{0}$ sample. The points with error bars are data; the solid line is the fit result; the blue dotted line is background, and the red area is the inclusive $D^{0}$ signal.}
\label{fg:m_rd} 
\end{figure}

The inclusive $D^{0}$ yield is extracted from a one-dimensional extended unbinned maximum likelihood fit, with the likelihood defined as 
\begin{equation}
\mathcal{L}=\frac{e^{-\sum_{j}N_{j}}}{N!}\prod^{N}_{i=1}(\sum_{j}N_{j}P_{j}(M_{D^{0}}^{i})),
\end{equation}
where $N$ is the total number of candidates, $N_{j}$ is the number of events in component $j$, $M_{D^{0}}^{i}$ is the $M_{D^{0}}$ value of the $i$th candidate, and $P_{j}$ represents the corresponding one-dimensional probability density function (PDF). There are two components in the fit: inclusive $D^{0}$ signal, modeled with a combination of two Gaussian functions and a bifurcated Gaussian function with common means, and the background, modeled with an ARGUS function~\cite{argus}. The free parameters in the fit are the yields of the two components and all the shape parameters except for the end-point of the ARGUS function, which is fixed by MC simulation. The fit is shown in Fig.~\ref{fg:m_rd}, and we obtain $694667^{+1494}_{-1563}$ inclusive $D^{0}$ decays.

Candidates for invisible $D^{0}$ decays are identified by requiring no remaining final-state particles associated with $\overline{D}^{0}_{\textrm{sig}}$. More precisely, events from the inclusive $\overline{D}^{0}_{\textrm{sig}}$ sample with remaining charged tracks, $\pi^{0}$, $K^{0}_{L}$, $K^{0}_{S}$, or $\Lambda$ are vetoed. In addition to $M_{D^{0}}$, the residual energy in the ECL, denoted as $E_{\textrm{ECL}}$, is also used to extract the $D^{0}\to$ invisible signal. The $E_{\textrm{ECL}}$ is defined as the sum of the energies of the ECL clusters that are not associated with the particles of the $D^{(*)}_{\textrm{tag}}X_{\textrm{frag}}\pi^{-}_{s}$ system. In order to suppress the beam background, cluster energies are required to be above ECL-region-dependent thresholds: 50 MeV for $32.2^{\circ}<\theta<128.7^{\circ}$, 100 MeV for $\theta<32.2^{\circ}$, and 150 MeV for $\theta>128.7^{\circ}$. 

We consider two backgrounds for the $D^{0}\to$ invisible signal: the $D^{0}$ background from the $e^{+}e^{-}\to c\overline{c}$ process in which correctly-tagged $D^{0}$ peak in $M_{D^{0}}$ (e.g. $D^0\to K^{0}\pi^{0}$) and the non-$D^{0}$ background from $e^{+}e^{-}\to q\overline{q}~(q=u,d,s,c)$, $\Upsilon(4S)$, and $\Upsilon(5S)$ decays. The signal yield is extracted from a two-dimensional extended unbinned maximum likelihood fit, with the likelihood defined as 
\begin{equation}
\mathcal{L}=\frac{e^{-\sum_{j}N_{j}}}{N!}\prod^{N}_{i=1}(\sum_{j}N_{j}P_{j}(M_{D^{0}}^{i},E_{\textrm{ECL}}^{i})),
\end{equation}
where $P_{j}$ represents the corresponding two-dimensional PDF, and $E_{\textrm{ECL}}^{i}$ is the $E_{\textrm{ECL}}$ value of the $i$th candidate. The $P_{j}$ functions are products of $M_{D^{0}}$ PDFs and $E_{\textrm{ECL}}$ PDFs since correlations between $M_{D^{0}}$ and $E_{\textrm{ECL}}$ are found to be small. There are three components in the fit: signal, $D^{0}$ background, and non-$D^{0}$ background. The PDFs in $E_{\textrm{ECL}}$ are histograms obtained from MC simulation. The $D^{0}$ and non-$D^{0}$ background PDFs in $E_{\textrm{ECL}}$ have a small peaking structure near $E_{\textrm{ECL}}=0$ GeV, and the corresponding systematic effects are described below. 
The signal PDF in $M_{D^{0}}$ is fixed as the one obtained by the fit to the $M_{D^{0}}$ distribution of the inclusive $D^{0}$ sample. The $D^0$ background PDFs in $M_{D^{0}}$ is parametrized with the sum of three Gaussian functions. The non-$D^0$ background PDF in $M_{D^{0}}$ is an ARGUS function. 
The free parameters in the fit are the yields of the three components, the $D^0$ background PDF shape parameters, and the non-$D^0$ background PDF shape parameters except for the end-point of the ARGUS function.

\begin{figure}[htb]
\centering
\includegraphics[width=0.4\textwidth]{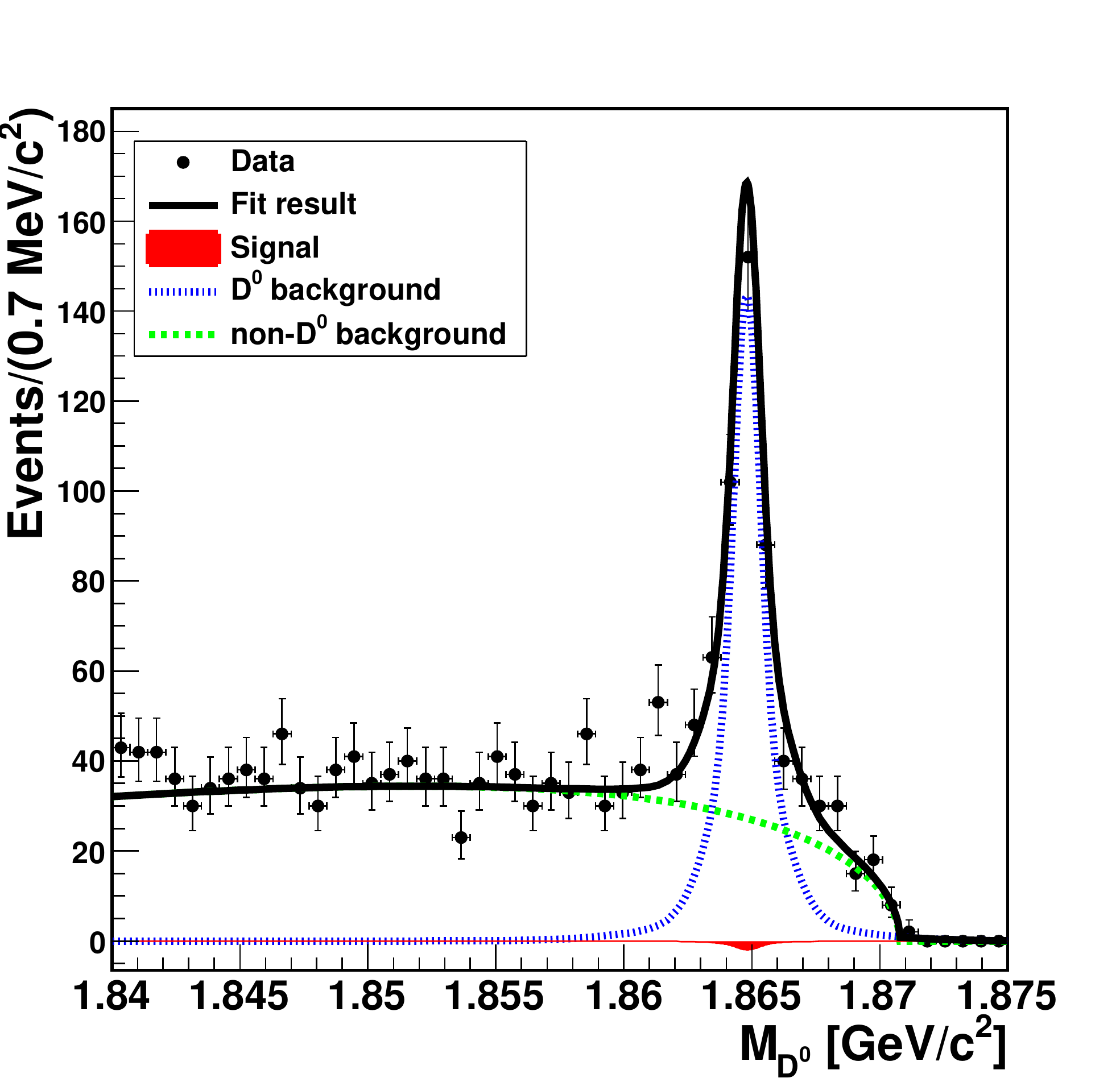}
\includegraphics[width=0.4\textwidth]{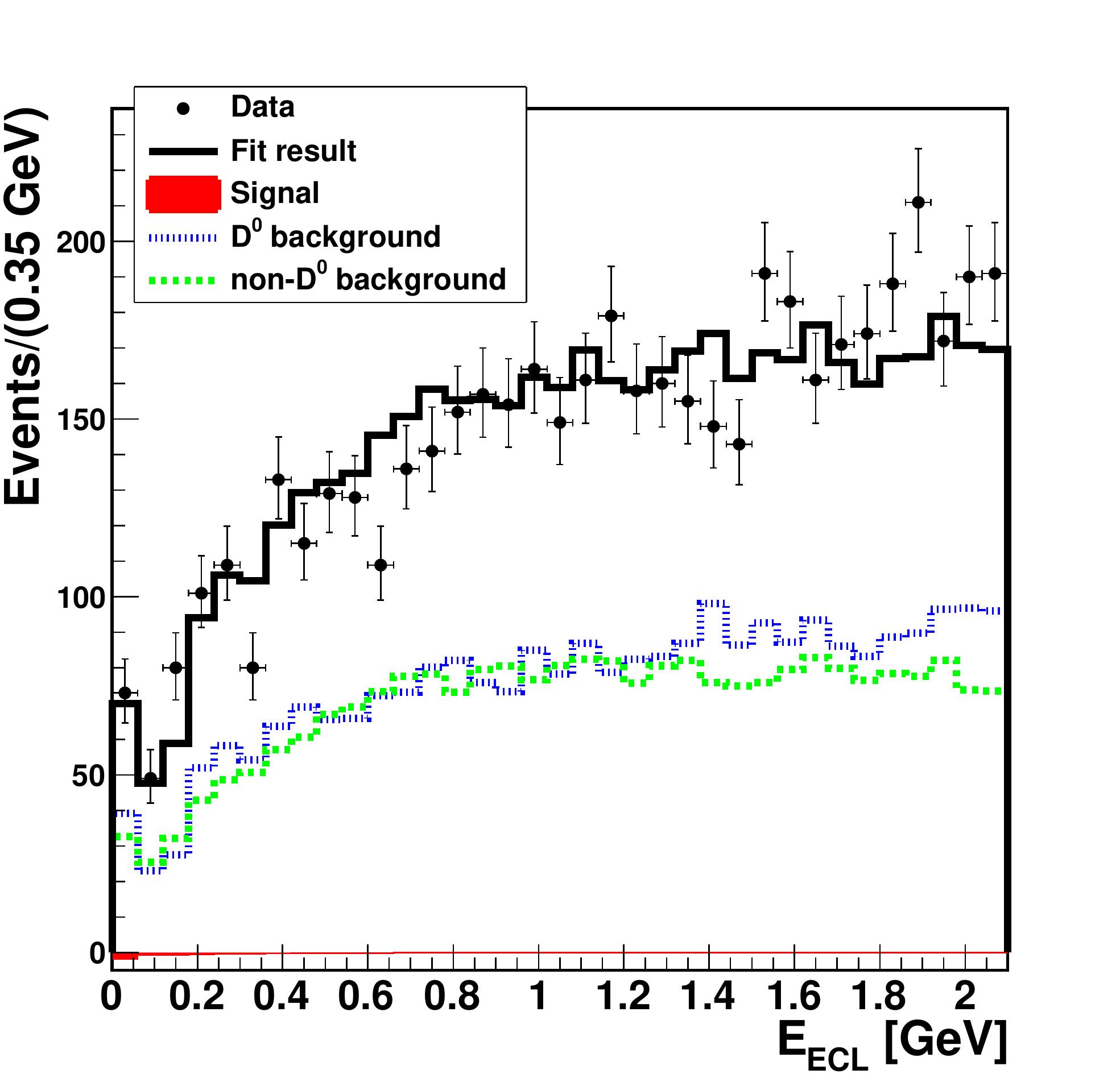}
\caption{Fit results of $D^{0}\to$ invisible decays. The top panel shows the $M_{D^{0}}$ distribution for $E_{\textrm{ECL}} <$ 0.5 GeV and the bottom one shows $E_{\textrm{ECL}}$ for $M_{D^{0}}$ $>$ 1.86 GeV/$c^{2}$.
The points with error bars are data; the solid line is the fit result; the blue dotted line is $D^{0}$ background; the green dashed line is non-$D^{0}$ background, and the red area is the signal of $D^{0}$ decaying to invisible final states.}
\label{fg:fit} 
\end{figure}

The projections of the fit are shown in Fig.~\ref{fg:fit}. 
The fitted signal yield of $D^{0}\to$ invisible is $-6.3^{+22.5}_{-21.0}$, which is consistent with zero. 

The branching fraction is calculated using 
\begin{equation}
\mathcal{B}=\frac{N_{\rm{sig}}}{\epsilon \times N^{\textrm{incl.}}_{D^{0}}},
\end{equation}
where $N_{\rm{sig}}$, $N^{\textrm{incl.}}_{D^{0}}$, and $\epsilon$ are the fitted signal yield of $D^{0}\to$ invisible decays, 
the number of inclusive $D^{0}$ mesons, and the efficiency of reconstructing $D^{0}\to$ invisible decays within the inclusive $D^{0}$ sample, respectively. 
We calibrate the reconstruction efficiency, estimated using the MC simulation by including in $\epsilon$ a factor $C_{\textrm{veto}}=1.1$ due to the corrections associated with the vetoes on the remaining final state particles in the reconstruction of $\overline{D}^{0}_{\textrm{sig}}$. The $C_{\textrm{veto}}$ value is obtained from a study with $D^{0}\to K^{-}\pi^{+}$ control sample described below. The calibrated reconstruction efficiency for the signal is $(62.4^{+3.2}_{-3.1})\%$.

\begin{figure}[htb]
\centering
\includegraphics[width=0.4\textwidth]{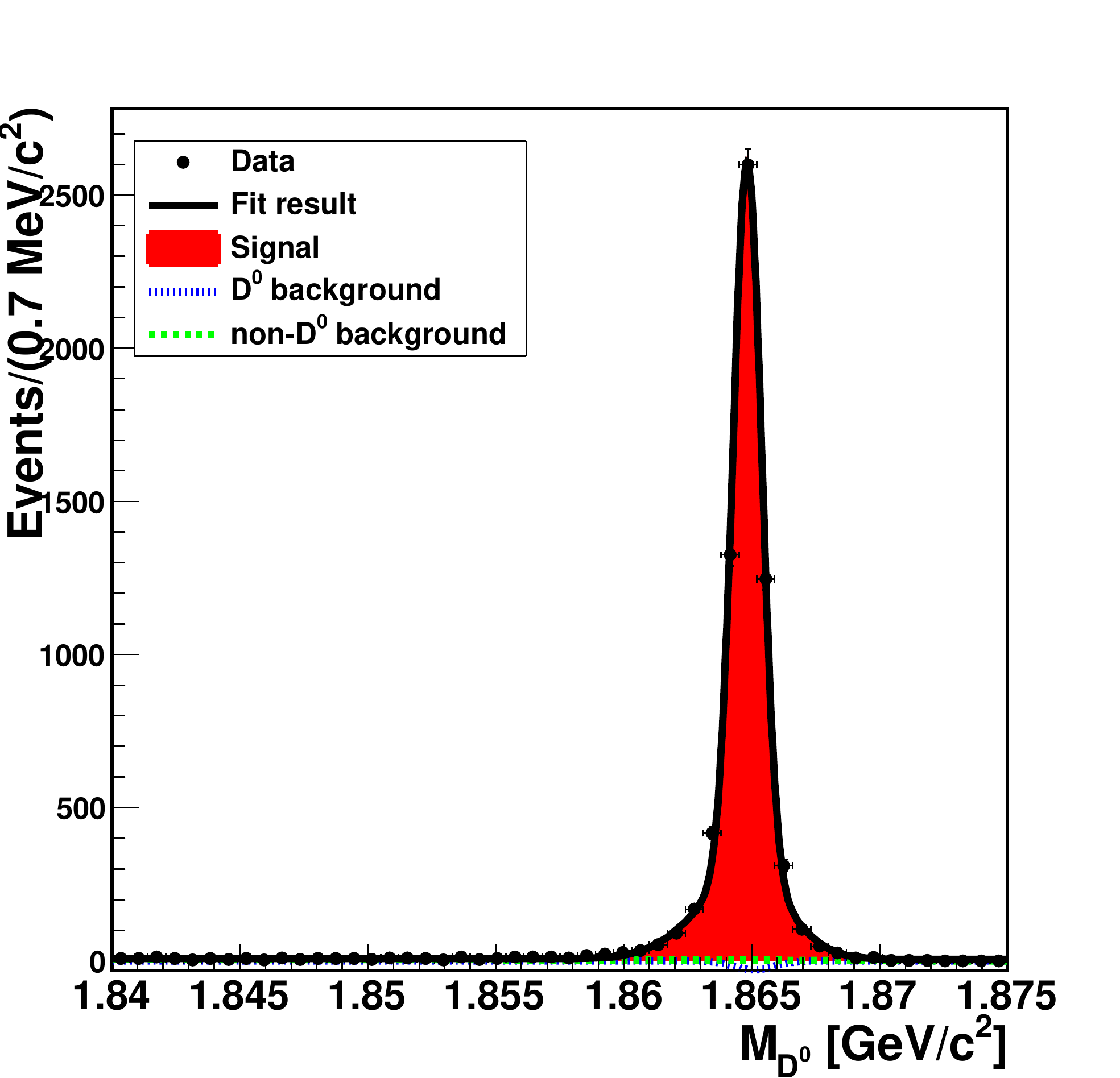}
\includegraphics[width=0.4\textwidth]{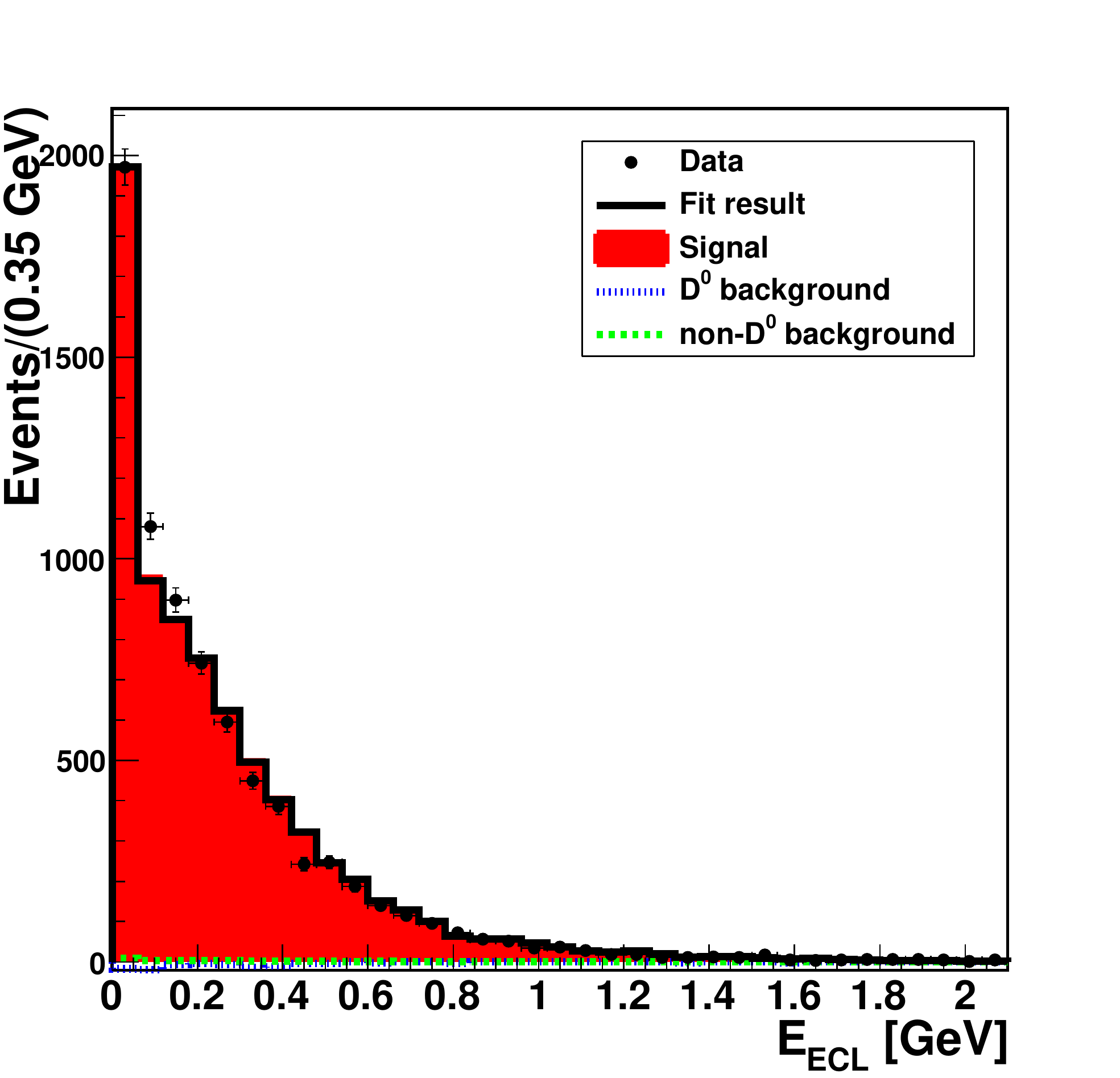}
\caption{Fit results of $D^{0}\to K^{-}\pi^{+}$. The top panel shows the $M_{D^{0}}$ distribution for $E_{\textrm{ECL}} <$ 0.5 GeV and the bottom one shows $E_{\textrm{ECL}}$ for $M_{D^{0}}$ $>$ 1.86 GeV/$c^{2}$.
The points with error bars are data; the solid line is the fit result; the blue dotted line is $D^{0}$ background; the green dashed line is non-$D^{0}$ background, and the red area is the $D^{0}\to K^{-}\pi^{+}$ signal.}
\label{fg:kpi_fit} 
\end{figure}

As a check, we repeat the entire analysis with the $D^{0}\to K^{-}\pi^{+}$ control sample.
After $D^{0}\to K^{-}\pi^{+}$ candidates are reconstructed from tracks associated with $\overline{D}^{0}_{\textrm{sig}}$ and $M_{K^{-}\pi^{+}}$ is required to be between 1.80 and 1.92 GeV/$c^{2}$, exactly the same selection criteria as for the $D^{0}\to$ invisible analysis are applied, excluding $K^{-}$ and $\pi^{+}$ from $\overline{D}^{0}_{\textrm{sig}}$. The fit result is shown in Fig.~\ref{fg:kpi_fit}. The efficiency of reconstructing $D^{0}\to K^{-}\pi^{+}$ is 29.0\%. With a signal yield of $7842^{+116}_{-117}$, we obtain $\mathcal{B}(D^{0}\to K^{-}\pi^{+})=$(3.89$\pm$0.06(stat.))\%, which is consistent with the world average of $(3.93\pm0.04)$\%~\cite{PDG}.

Sources of various systematic uncertainties on the branching fraction calculation are shown in Table~\ref{tb:sys}. The uncertainties associated with $\epsilon$ and $N^{\textrm{incl.}}_{D^{0}}$ are quoted as percentages, while the uncertainties associated with signal yield extraction are quoted as event yields. The uncertainty due to the yield of inclusive signal $D^{0}$ mesons includes the statistical and systematic uncertainties. The latter includes uncertainties due to signal $D^{0}$ PDF and background PDF modeling, and these are obtained by the variation of the measured yield using different shape functions in the $D^{0}\to K^{-}\pi^{+}$ fit and the fit to the inclusive $D^{0}$ mass spectrum, respectively. The calibration factor $C_{\textrm{veto}}$ and the associated systematic uncertainty are obtained by comparing the data ($\epsilon_{\textrm{data}}$) and MC veto efficiency ($\epsilon_{\textrm{MC}}$) using the $D^{0}\to K^{-}\pi^{+}$ control sample. In addition, the ratios $\epsilon_{\textrm{data}}/\epsilon_{\textrm{MC}}$ with different $D^{(*)}_{\textrm{tag}}$/$X_{\textrm{frag}}$ reconstruction modes are studied and are found to be consistent with each other within $\pm1\sigma$ of their statistical uncertainty; the variation is included in the systematic uncertainty. The statistical uncertainty of the MC sample in the efficiency estimation is also included.

No contribution to systematic uncertainty is expected from the uncertainties of the $M_{D^{0}}$ PDF parameters of the $D^{0}$ background as they are free in the fit.  However, possible imperfection of functional form and the correlation between $M_{D^{0}}$ and $E_{\textrm{ECL}}$ PDFs may cause systematic bias in the signal yield. The uncertainty due to such a possible yield bias is estimated by an MC ensemble test with an assumed branching fraction of zero. The uncertainties due to the shape-fixed PDF in the fit are obtained from the signal yield change when varying the PDF shape. For the signal PDF in $E_{\textrm{ECL}}$, the histogram PDF is varied by the data-MC difference in the $E_{\textrm{ECL}}$ distribution of the $D^{0}\to K^{-}\pi^{+}$ control sample. For the $D^{0}$ background PDF in $E_{\textrm{ECL}}$, we vary the first-bin content of the histograms by $\pm1\sigma$ of the branching fraction of the $D^{0}$ decay modes, where $\sigma$ denotes the measurement error on the branching fraction. For the non-$D^{0}$ background PDF in $E_{\textrm{ECL}}$, we find that the MC can describe data well in the region $M_{D^{0}}$ $<$ 1.855 GeV/$c^{2}$, and the histogram PDF is also varied by the data-MC difference in the $E_{\textrm{ECL}}$ distribution in this region. For the signal PDF in $M_{D^{0}}$, we vary the shape parameters by $\pm1\sigma$, where $\sigma$ denotes standard deviation of the shape parameters obtained by the fit on $M_{D^{0}}$ distribution of the inclusive $D^{0}$ sample. For the non-$D^{0}$ background PDF in $M_{D^{0}}$, we float the end-point in the fit and the signal yield variation is found to be negligible.

\begin{table}[htbp]
\begin{center}
\caption{Summary of the systematic uncertainties on the branching fraction.}
\begin{tabular}{c|c}
\hline
\hline
Source & In \% \\ \hline
$N^{\textrm{incl.}}_{D^{0}}$ & $\pm0.2$(stat.) $\pm3.6$(syst.) \\ 
$C_{\textrm{veto}}$ & $+4.7$/ $-4.6$ \\ 
MC statistics & $\pm1.9$ \\ \hline 
Total & $+6.2$/ $-6.1$ \\ \hline \hline
\multicolumn{2}{c}{}  \\ \hline \hline
Source & In events \\ \hline
Yield bias & $-0.5$ \\
Signal PDF in $E_{\textrm{ECL}}$ & $+2.3$ \\
$D^{0}$ background PDF in $E_{\textrm{ECL}}$ & $+2.5$/ $-2.6$ \\
Non-$D^{0}$ background PDF in $E_{\textrm{ECL}}$ & $-13.7$ \\
Signal PDF in $M_{D^{0}}$ & $+0.2$/ $-0.4$ \\ 
Non-$D^{0}$ background PDF in $M_{D^{0}}$ & negligible \\ \hline
Total & $+3.4$/ $-14.0$ \\ \hline \hline
\end{tabular}
\label{tb:sys}
\end{center}
\end{table}

Since the observed yield for $D^{0}\to$ invisible is not significant, we calculate a
90$\%$ confidence level Bayesian upper limit on the branching fraction ($\mathcal{B}_{\textit{UL}}$)~\cite{UL}. The upper limit is obtained by integrating the likelihood function:
\begin{equation}
\int^{\mathcal{B}_{\textit{UL}}}_{0}\mathcal{L}(\mathcal{B})d\mathcal{B} = 0.9\int^{1}_{0}\mathcal{L}(\mathcal{B})d\mathcal{B},
\end{equation}
where $\mathcal{L}(\mathcal{B})$ denotes the likelihood value.
The systematic uncertainties are taken into account by replacing $\mathcal{L}(\mathcal{B})$ with a 
smeared likelihood function:
\begin{equation}
\mathcal{L}_{\textrm{smear}}(\mathcal{B}) = \int^{1}_{0} \mathcal{L}(\mathcal{B}') \frac{e^{-\frac{(\mathcal{B}-\mathcal{B}')^{2}}{2\Delta \mathcal{B}^{2}}}}{\sqrt{2\pi}\Delta \mathcal{B}}d\mathcal{B}',
\end{equation}
where $\Delta \mathcal{B}$ is the total systematic uncertainty on $\mathcal{B}'$.
We thus determine the upper limit on the branching fraction of $D^{0} \to$ invisible to be $9.4 \times 10^{-5}$ at the 90\% confidence level.

In conclusion, we have performed the first search for $D^{0}$ decays into invisible final states with the charm tagger method by using a data sample of 924 fb$^{-1}$ collected by Belle. No significant signal yield is found and we set an upper limit on the branching fraction of $9.4\times 10^{-5}$ at the 90\% confidence level for the $D^{0}\to$ invisible decay. Further improvement in this measurement may be possible in the near future with other $e^{+}e^{-}$ collider experiments such as BESIII and Belle II.

We thank the KEKB group for the excellent operation of the
accelerator; the KEK cryogenics group for the efficient
operation of the solenoid; the KEK computer group,
the National Institute of Informatics, and the 
PNNL/EMSL computing group for valuable computing
and SINET4 network support.  We acknowledge support from
the Ministry of Education, Culture, Sports, Science, and
Technology (MEXT) of Japan, the Japan Society for the 
Promotion of Science (JSPS), and the Tau-Lepton Physics 
Research Center of Nagoya University; 
the Australian Research Council;
Austrian Science Fund under Grants No.~P 22742-N16 and No.~P 26794-N20;
the National Natural Science Foundation of China under Contracts 
No.~10575109, No.~10775142, No.~10875115, No.~11175187, No.~11475187
and No.~11575017;
the Chinese Academy of Science Center for Excellence in Particle Physics; 
the Ministry of Education, Youth and Sports of the Czech
Republic under Contract No.~LG14034;
the Carl Zeiss Foundation, the Deutsche Forschungsgemeinschaft, the
Excellence Cluster Universe, and the VolkswagenStiftung;
the Department of Science and Technology of India; 
the Istituto Nazionale di Fisica Nucleare of Italy; 
the WCU program of the Ministry of Education, National Research Foundation (NRF) 
of Korea Grants No.~2011-0029457,  No.~2012-0008143,  
No.~2012R1A1A2008330, No.~2013R1A1A3007772, No.~2014R1A2A2A01005286, 
No.~2014R1A2A2A01002734, No.~2015R1A2A2A01003280 , No. 2015H1A2A1033649;
the Basic Research Lab program under NRF Grant No.~KRF-2011-0020333,
Center for Korean J-PARC Users, Grant No.~NRF-2013K1A3A7A06056592; 
the Brain Korea 21-Plus program and Radiation Science Research Institute;
the Polish Ministry of Science and Higher Education and 
the National Science Center;
the Ministry of Education and Science of the Russian Federation and
the Russian Foundation for Basic Research;
the Slovenian Research Agency;
Ikerbasque, Basque Foundation for Science and
the Euskal Herriko Unibertsitatea (UPV/EHU) under program UFI 11/55 (Spain);
the Swiss National Science Foundation; 
the Ministry of Education and the Ministry of Science and Technology of Taiwan;
and the U.S.\ Department of Energy and the National Science Foundation.
This work is supported by a Grant-in-Aid from MEXT for 
Science Research in a Priority Area (``New Development of 
Flavor Physics'') and from JSPS for Creative Scientific 
Research (``Evolution of Tau-lepton Physics'').


\begin{thebibliography}{99}
\bibitem{charge_conjugate} Throughout this paper, inclusion of charge-conjugate decay modes is always implied.
\bibitem{UL} As we use a Bayesian method, this is formally a ``credibility level." However, we use ``confidence level" here following common convention.
\end{thebibliography}
\end{document}